\documentstyle[epsf]{europhys}
\def\etal{{\hbox{{\tenit\ et al.\/}\tenrm :\ }}}

\def\And{{\rm and\ }}

\def\drm{{\rm d}}

\def\dif#1#2{\frac{\drm#1}{\drm#2}}

\def\order{\mathop{\rm O}\nolimits}

\def\stars{\bigskip\centerline{***}\medskip}

\newif\ifboo \boofalse

\def\Review#1{\boofalse{\it #1},}
\def\Name#1{{\sc #1},}
\def\Vol#1{\ifboo Vol. {\bf #1}\else{\bf #1}\fi}
\def\Year#1{\ifboo #1\else(#1)\fi}
\def\Book#1{\bootrue{\it #1},}
\def\Page#1{\ifboo {\rm p. #1}\else{\rm #1}\fi}
\newcommand{\cD} {{\cal D}}
\newcommand{\cH} {{\cal H}}
\newcommand{\eg} {{\it e.g., }}
\newcommand{\ex} {{\rm e}}
\newcommand{\ie} {{\it i.e., }}
\newcommand{\talpha} {\tilde{\alpha}}
\newcommand{\tlambda} {\tilde{\lambda}}
\newcommand{\vu} {{\bf u}}
\def\trace{\mathop{\mbox{\large Tr}}}

\begin{document}
\euro{}{}{}{}
\Date{}
\shorttitle{H. DIAMANT \etal SELF-ASSEMBLY OF AMPHIPHILIC POLYMERS}
\title{Self-assembly in mixtures of amphiphilic polymers and\\ 
    surfactants}
\author{H. Diamant \And D. Andelman}
\institute{School of Physics and Astronomy\\
    Raymond and Beverly Sackler Faculty of Exact Sciences\\
    Tel Aviv University, 69978 Tel Aviv, Israel}
\rec{}{}
\pacs{
\Pacs{61}{25Hq}{Macromolecular and polymer solutions, polymer melts, 
    swelling}
\Pacs{61}{41+e}{Polymers, elastomers, and plastics}
\Pacs{87}{15Da}{Physical chemistry of solutions of biomolecules,
    condensed states}
      }
\maketitle
%
%%%   The Abstract
%
\begin{abstract}
We present a model for the joint 
self-assembly of amphiphilic polymers and small 
amphiphilic molecules (surfactants) in a dilute aqueous 
solution.
The polymer is assumed to consist of a hydrophilic 
backbone and a large number of hydrophobic side groups.
Preference of the surfactant to bind to hydrophobic
microdomains along the polymer induces an effective attraction
between bound surfactants.
This leads to two distinct binding regimes
depending on a single physical parameter, $\epsilon$, which 
represents the ratio between surfactant-polymer affinity and 
polymer hydrophobicity.
For small $\epsilon$ the binding is non-cooperative,
whereas for large $\epsilon$ it becomes strongly cooperative
at a well-defined critical aggregation concentration.
Our findings are in accord with observations on diverse
experimental systems.
\end{abstract}
%
%%%   Main text
%
A considerable experimental and theoretical effort has been
devoted in recent years to the study of mixtures of
polymers and amphiphilic molecules (surfactants) in solution 
\cite{review}.
Of particular interest and importance 
are systems where the polymer itself has
amphiphilic qualities 
as they exhibit rich self-assembly behaviour \cite{hmp}.
These systems offer, on one hand, numerous industrial
applications, since their properties (\eg rheology) can be 
conveniently tuned and their solubility in water
makes them ``environmentally friendly''. 
On the other hand, their complex amphiphilic nature poses
a challenge to current theoretical investigations
far beyond more conventional polymers in solution.
Besides industrial applications,
biological macromolecules, such as proteins, RNA and
single strands of DNA,
may be viewed as special, more complex cases of 
amphiphilic polymers.

\begin{figure}
\vbox to 6cm{\vfill\centerline{
\epsfxsize=0.4\linewidth
\epsfysize=0.4\linewidth
\fbox{\epsffile{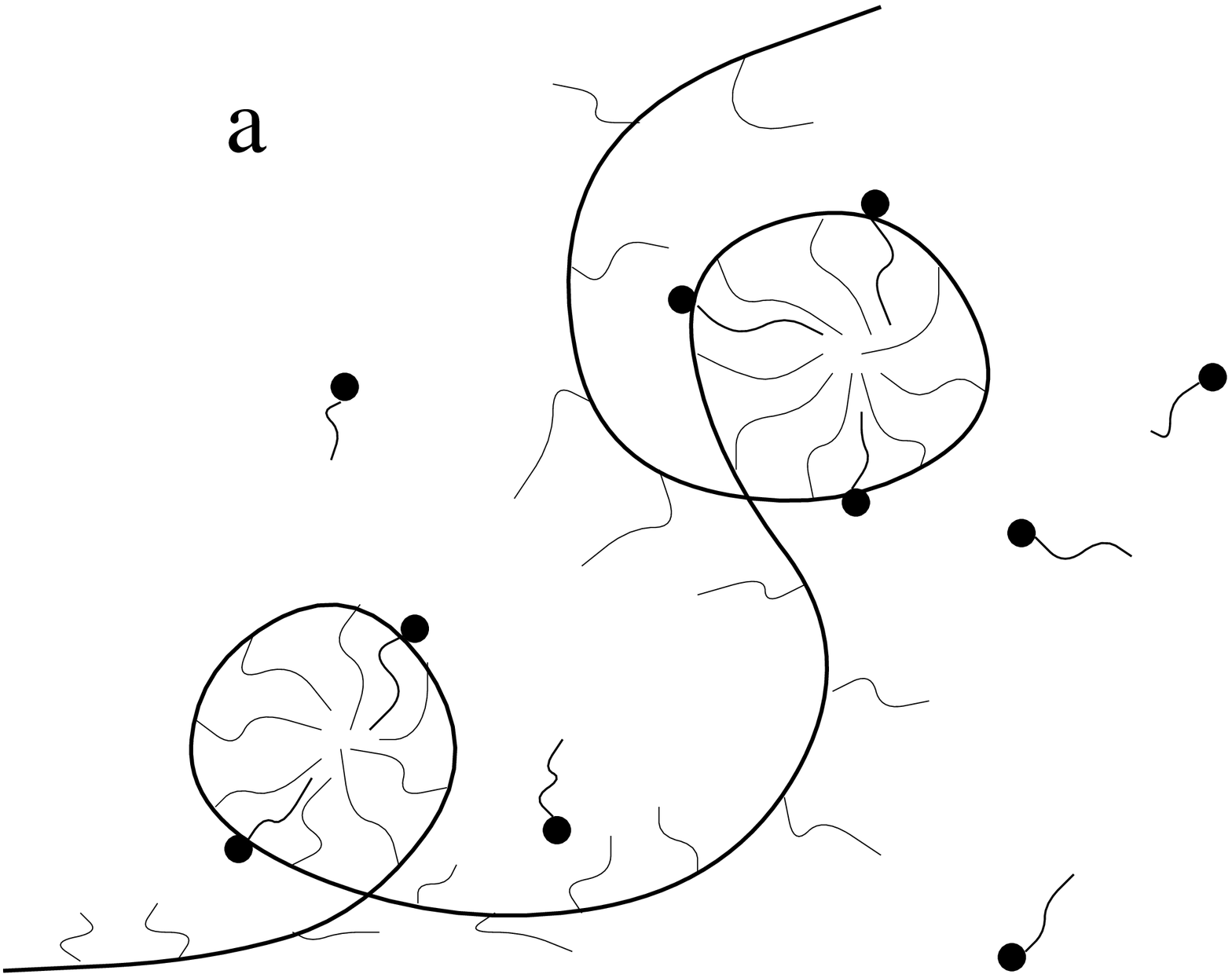}}
\epsfxsize=0.4\linewidth
\epsfysize=0.4\linewidth
\fbox{\epsffile{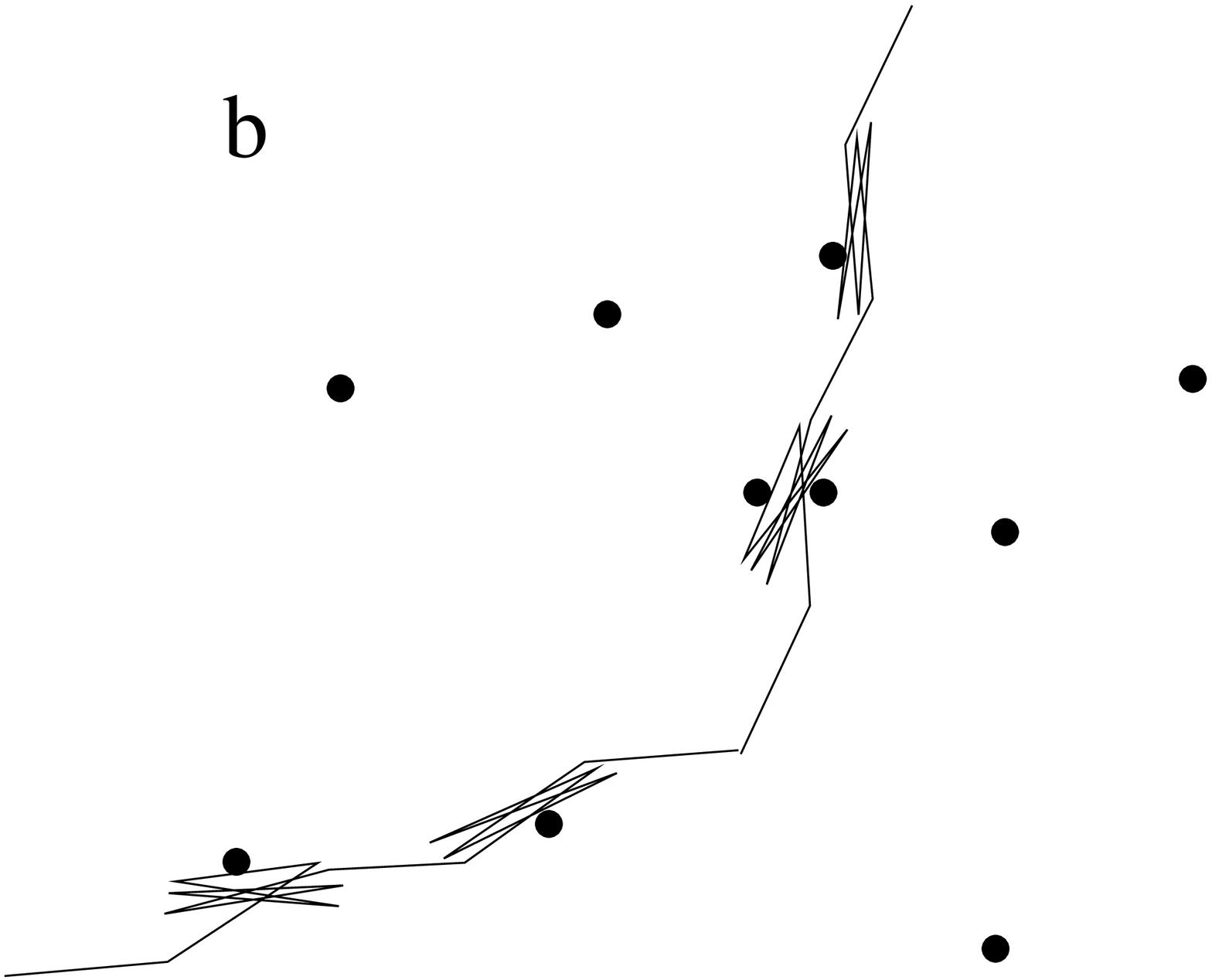}}}\vfill}
\caption[]
{ (a) Schematic view of a section of a side-chain 
   amphiphilic polymer with bound surfactants.
   (b) Schematic view of the system as described by our model.} 
\label{scheme}
\end{figure}

In the current work we focus on the case where a large 
fraction of hydrophobic side groups are chemically 
attached to a hydrophilic polymer backbone 
(fig.~\ref{scheme}a).
The properties of such macromolecules and their
interaction with free surfactants in solution have been 
investigated in a series of experiments during the past 
years, revealing rich self-assembly behaviour
\cite{Zana,Talmon}.
One of the notable findings is that the cooperativity of the 
surfactant-polymer binding, which is related to interactions
between bound surfactants, sensitively depends on the 
polymer properties (\eg its hydrophobic side chains and the 
degree of ionisation of its backbone).
Previous theories treating polymer-surfactant mixtures
\cite{phenomen,micelle,polysoap} did not address the 
diverse behaviour of those experimental systems.
Our aim in this Letter is to construct a simple model
which yields a qualitative understanding of the
thermodynamics of such systems while accounting, 
in particular, for the polymer-induced interactions between
bound surfactants.
We neglect, therefore, structural details such as
the size of the hydrophobic side chains, steric effects, 
chain rigidity etc.,
which are microscopically important but not
expected to qualitatively affect the thermodynamics.
A more refined model, including structural details,
has allowed us to verify this assumption and will
be presented in a future paper \cite{future}.
Transmission Electron Micrographs of dilute 
solutions of side-chain amphiphilic polymers, with and
without surfactant, show self-assembled hydrophobic 
microdomains forming locally along the polymeric chains,
while the chains seem to maintain overall 
extended conformations \cite{Talmon}.
This observation is explained by the highly hydrophilic
polymer backbone (usually a polyelectrolyte) and the
high density of hydrophobic side groups along the chain.
We are allowed, therefore, to consider only local
association along the chain sequence, which greatly
simplifies the model.
The current work is restricted to the dilute limit, 
where inter-chain effects can be neglected and the 
surfactant is below its critical micelle concentration,
in accordance with the relevant experiments.

Consider a flexible chain of $N$ segments
in contact with a surfactant reservoir of chemical
potential $\mu_{\rm s}$.
To each segment we assign a vector, $\vu_n$, and
denote the number of surfactants bound to it by $\varphi_n$.
The energy gained by a segment joining a microdomain
is denoted by $\alpha$, and the one gained by
a surfactant bound to such a microdomain is 
$\gamma$. 
The energetic parameters, $\alpha$ and $\gamma$,
represent, respectively, measures of the polymer 
hydrophobicity and polymer-surfactant affinity.
(All energies in this Letter are in units of $k_{\rm B}T$.)
The partition function for the polymer-surfactant system
is
\begin{eqnarray}
\label{Zps1}
    Z_{\rm ps} &=& \trace_{\{\varphi_n\}} \int \cD\vu_n \exp( 
    -\cH_{\rm ps}/k_{\rm B}T)  \nonumber \\
    \cH_{\rm ps}/k_{\rm B}T &=&  \sum_{n=1}^N \lambda_n u_n^2  
    - \sum_{n=1}^{N-1} \frac{\alpha+\gamma\varphi_n}{4} 
      (\vu_{n+1}-\vu_n)^2 - \mu_{\rm s} \sum_{n=1}^N \varphi_n
\end{eqnarray}
where the $\lambda_n$ are Lagrange multipliers ensuring the
constraints $\langle u_n^2\rangle=1$ ($\langle\cdots\rangle$ denotes 
a thermal average over all chain configurations).
Some of our results can be more easily derived 
imposing infinitely stiff segment constraints ($u_n^2=1$).
Yet, other results, in particular the effect of polymer configurations
on surfactant binding, cannot be 
obtained as easily using such an approach.
We have modelled the association property of the polymer by
a local tendency of the segments to fold, and the preference
of the surfactant to join such clusters by a linear coupling to
this folding \cite{membcurv}.
Not explicitly included in eq.~(\ref{Zps1}), interactions
between surfactants may arise only
indirectly through the coupling to chain configurations.
Note that the current model contains no structural parameters.
The hydrophobic microdomains are described as featureless clusters
of folded polymer segments attracting free surfactants, as illustrated
in fig.~\ref{scheme}b.
In a more refined model, steric and other repulsive effects should compete 
with the association, resulting in a finite radius of curvature for
the microdomains \cite{future}.
Such microscopic details, however, do not alter the thermodynamic
behaviour which concerns us in this work.

In the surfactant-free case,
the partition function (\ref{Zps1}) becomes similar to the one for 
semi-flexible polymers \cite{HT},
\begin{equation}
  Z_{\rm p} = \int \cD\vu_n \exp[ -\sum_{n=1}^N \lambda_n u_n^2 
    + \frac{\alpha}{4} \sum_{n=1}^{N-1} (\vu_{n+1}-\vu_n)^2 ]
\label{Zp1}
\end{equation}
Yet, because of the tendency of segments in our
model to {\it anti}-align ($\alpha>0$), we need first to integrate 
over the ``sub-lattice'' $\vu_2,\vu_4,\ldots$ before taking the 
continuum limit.
In addition, we replace all $\lambda_n$ with
a single $\lambda$.
(This approximation amounts to a non-extensive 
correction in the free energy, which becomes negligible for 
$N\rightarrow\infty$ \cite{future}.)
The resulting expression is
\begin{equation}
\label{Zp2}
  Z_{\rm p} = \left(\frac{\pi}{\lambda-\alpha/2}\right)^{3N/4}
    \int \cD\vu(n) \exp [ -\frac{\talpha}{2} \int_0^N 
     \left(\dif{\vu}{n} \right)^2 \drm n
     -  \frac{\tlambda}{2} \int_0^N u^2(n) \drm n ]
\end{equation}
where $\tlambda \equiv \lambda(\lambda-\alpha)/(\lambda-\alpha/2)$,
$\talpha \equiv \alpha^2/4(\lambda-\alpha/2)$,
and $n: 0\rightarrow N$ is now a continuous parameter along the chain.
(Note that the continuum limit implies a large $\talpha$ and, hence,
a large $\alpha$.)
The path integral in eq.~(\ref{Zp2}) is readily evaluated using an 
analogy with a three-dimensional quantum oscillator \cite{Feynman} 
and the ground-state dominance \cite{Edwards}, yielding
\begin{equation}
    Z_{\rm p} = \left(\frac{\pi}{\lambda-\alpha/2}\right)^{3N/4}
        \exp\left[-(3N/2)\sqrt{\tlambda/\talpha}\right]
\end{equation}
Imposing now the constraint on $\langle u_n^2\rangle$,
$-\partial\log Z_{\rm p}/\partial\lambda = N$,
we find
    $\lambda=\alpha+\order(1/\alpha)$,
    $\tlambda=9/2\alpha+\order(1/\alpha^2)$,
    and $\talpha=\alpha/2+\order(1/\alpha)$.
The correlation between alternating segment directions 
along the chain is found from eq.~(\ref{Zp2}) as
\begin{equation}
  \langle\vu(n)\cdot\vu(n')\rangle = \exp(-|n-n'|/\xi), 
  \ \ \xi = \alpha/3
\end{equation}
The value of $\xi$ defines a chain length over which 
alternating segment directions, $\vu_n,\vu_{n+2}\ldots$, are
correlated. 
Since the other segments within this length,
$\vu_{n+1},\vu_{n+3}\ldots$, are assumed, on average,
to be anti-aligned with those vectors,
we identify $\xi$ as the average number of segments 
in a single folded cluster (see fig.~\ref{scheme}b),
\ie in a single hydrophobic microdomain.
Thus, only when $\alpha$ is significantly large does the polymer tend
to form microdomains by itself.
The free energy of the chain is
\begin{equation}
\label{Fp}
    F_{\rm p}(\alpha) = -\ln Z_{\rm p} - \lambda N
    = N [ -\alpha + ({3}/{4})\log({\alpha}/{2\pi})]
\end{equation}
and can be viewed as composed of competing 
curvature energy (decreasing with $\alpha$) and entropy (increasing with 
$\alpha$) terms.

\begin{figure}
\vbox to 7cm{\vfill\centerline{
\epsfxsize=0.5\linewidth
\epsfysize=0.5\linewidth
\hbox{\epsffile{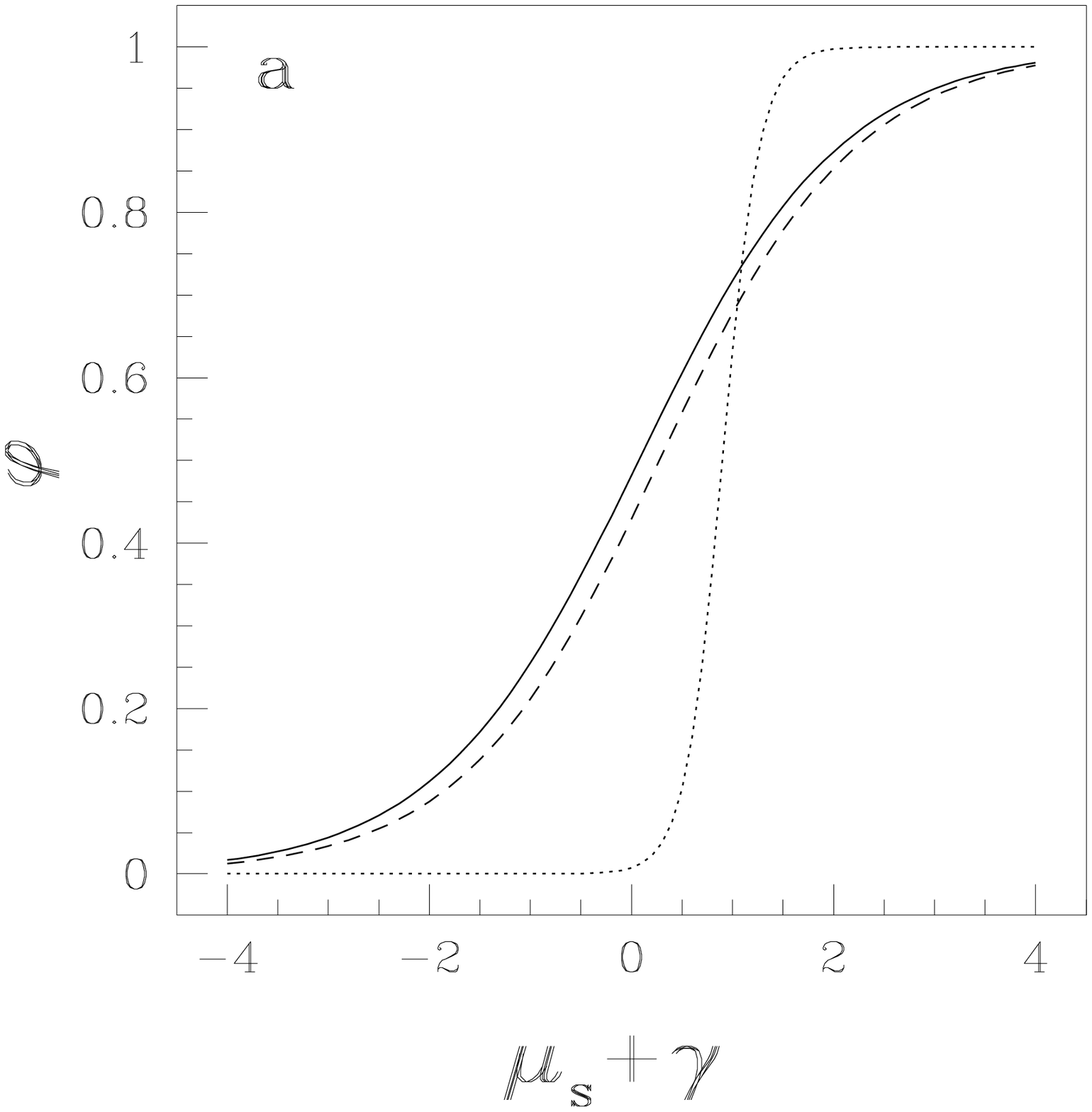}}
\epsfxsize=0.5\linewidth
\epsfysize=0.5\linewidth
\hbox{\epsffile{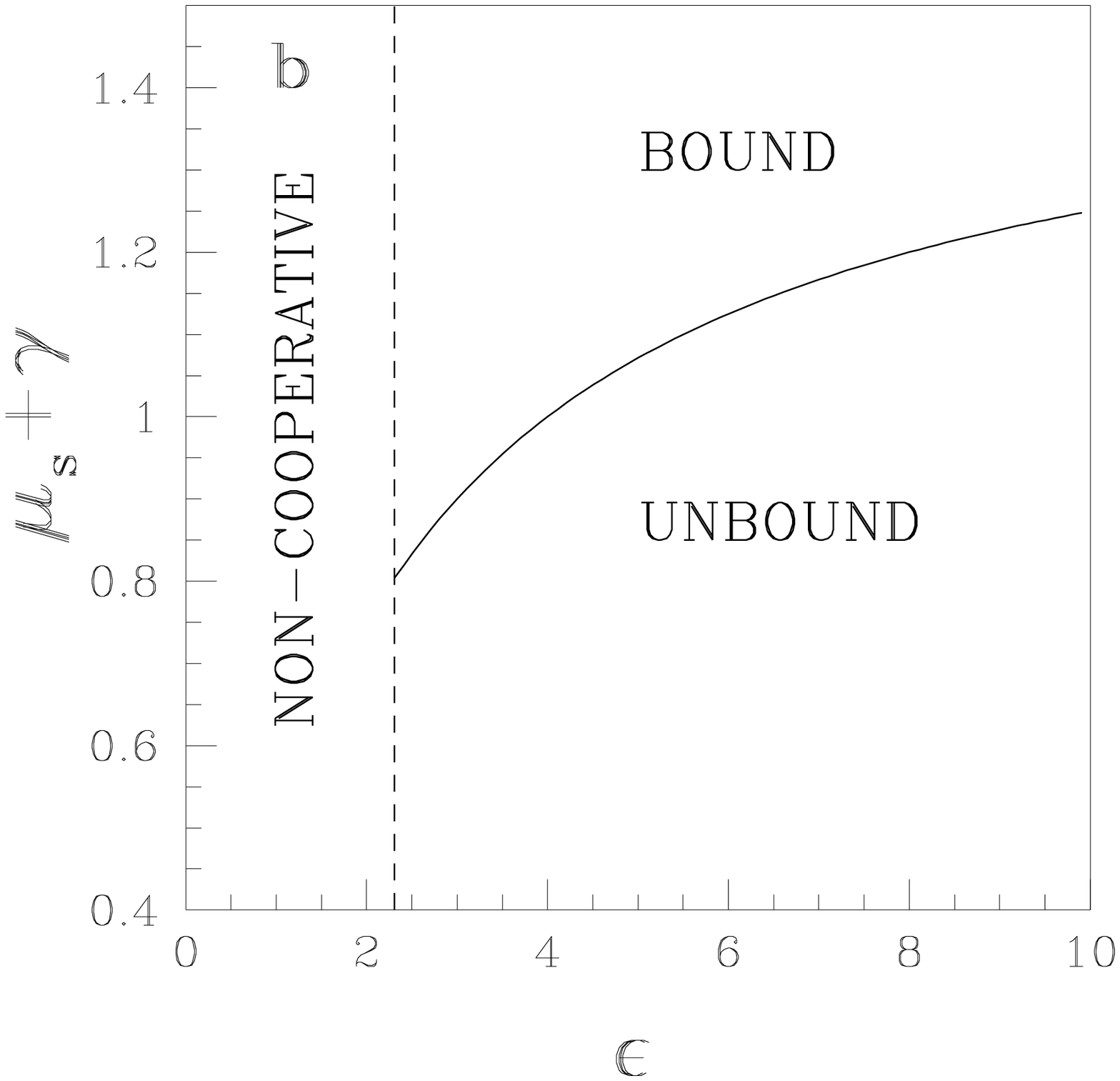}}}\vfill}
\caption[]
{ (a) Surfactant-polymer binding isotherms. Non-cooperative binding is 
   shown for $\epsilon=0.1$ (solid)
   and $0.5$ (dashed) using eq.~(\ref{isotherm});
   cooperative  binding is presented for
   $\epsilon=3$ (dotted) using eq.~(\ref{isotherm_all}).
   (b) Diagram presenting the surfactant-polymer binding behaviour. 
   The dashed line separates two binding regimes:
   for small $\epsilon$ the binding is non-cooperative, whereas
   for large $\epsilon$ the binding is cooperative and exhibits a sharp
   (though continuous) increase at the {\it cac}. The values of 
   $\mu_{\rm s}+\gamma$ corresponding to the {\it cac} are 
   represented by the solid line.}
\label{results}
\end{figure}

We now return to the full polymer-surfactant partition 
function (\ref{Zps1}).
For simplicity let us assume that at most one surfactant
can bind to a monomer, \ie $\varphi_n$ is either 0 or 1.
Tracing over $\{\varphi_n\}$ and expanding to
2nd order in $\epsilon\equiv\gamma/\alpha$, we find the following 
expression for the free energy:
\begin{equation}
\label{Fps}
    \frac{F_{\rm ps}}{N} \simeq \frac{F_{\rm p}}{N}
    + \ln(1-f) + \frac{3}{4} f \epsilon
    - \frac{3}{32} f(3+f) \epsilon^2
\end{equation}
where $f \equiv [1+\ex^{-(\mu_{\rm s}+\gamma)}]^{-1}$
is the average fraction of bound surfactant if
it had been simply attracted to the polymer without
coupling to its curvature.
The first term in eq.~(\ref{Fps}) is the free energy 
of the surfactant-free chain, eq.~(\ref{Fp}),
the second describes a regular lattice-gas contribution,
and the remaining terms represent the surfactant-polymer
interaction. 
From eq.~(\ref{Fps}) a binding isotherm can be calculated,
\begin{equation}
\label{isotherm}
   \varphi \simeq f [ 1 - (3/4)(1-f)\epsilon
   + (3/32)(1-f)(3+2f)\epsilon^2 ]
\end{equation}
where $\varphi$ is the average fraction of bound surfactant.
Figure~\ref{results}a shows two such isotherms (solid and dashed lines) for two 
different, small values of $\epsilon$.
In this regime the binding increases gradually with
$\mu_{\rm s}$ (non-cooperative binding).
For larger values of $\epsilon$ the isotherms are shifted to larger 
$\mu_{\rm s}$ and become steeper.
Let us define a chemical potential, 
$\mu_{\rm s}^*$, for which the amount of bound surfactant
is appreciable, \ie $\varphi=1/2$.  A corresponding 
cooperativity parameter is defined as
$C\equiv\partial\varphi/\partial\mu_{\rm s}|_{\mu_{\rm s}^*}$.
From eq.~(\ref{isotherm}) we readily find
\begin{eqnarray}
 \label{cac_small}
    \mu_{\rm s}^* &\simeq& -\gamma+(3/4)\epsilon-(3/8)\epsilon^2 \\
 \label{C_small}
    C &\simeq& (1/4)[1+(3/16)\epsilon^2]
\end{eqnarray}
The last, positive contributions to the average 
binding (\ref{isotherm}) and cooperativity (\ref{C_small})
indicate an effective short-range attraction between 
bound surfactants induced by chain configurations.
Once a molecule has bound to the chain, a folded hydrophobic 
region is formed, favouring binding of another molecule  
to the same region.
Effective attraction is a general feature of annealed impurities.
The important point, however, is that the attraction 
strength (resulting in binding cooperativity) depends on a single 
parameter, $\epsilon$, denoting the ratio between 
surfactant-polymer affinity and polymer hydrophobicity.

For larger $\epsilon$, the value of $\mu_{\rm s}^*$ can be 
estimated as
\begin{equation}
\label{cac_all}
   \mu_{\rm s}^* = \left. -\frac{\gamma}{4}\langle(\vu_{n+1}-\vu_n)^2
   \rangle \right|_{\varphi=1/2}
   = -\gamma + \frac{3\epsilon}{4(1+\epsilon/2)}
\end{equation}
To obtain similar estimates for the binding isotherm and cooperativity
we return to eq.~(\ref{Zps1}) and make a cumulant
expansion about the surfactant-free Hamiltonian {\it before} tracing 
over $\{\varphi_n\}$.
Restricting to two-body interactions between nearest bound surfactants, 
the resulting partition function becomes analogous to
a one-dimensional lattice-gas or Ising model with an attraction term 
proportional to $\epsilon^2$.
\footnote{For large $\epsilon$, many-body interactions among 
surfactants may become significant. Yet, since in our case all such
contributions are attractive, they would merely enhance the binding 
cooperativity and further support our findings.}
Calculation of the binding isotherm and cooperativity is then straightforward
and gives
\begin{eqnarray}
\label{isotherm_all}
  \varphi &=& \frac {g[g-1+\sqrt{(g-1)^2+2h}]+h}
  {(g-1)^2+(g+1)\sqrt{(g-1)^2+2h}+2h}; \ \ 
  g \equiv \frac{f}{1-f}\ex^{-3\epsilon/4}, \ \ 
  h \equiv 2g\ex^{-3\epsilon^2/8} \\ 
\label{C_all}
   C &=&  (1/4)\exp[(3/16)\epsilon^2]
\end{eqnarray}
For small $\epsilon$, eqs.~(\ref{cac_all})--(\ref{C_all}) 
coincide with eqs.~(\ref{isotherm})--(\ref{C_small}).
Equation (\ref{C_all}) is an important result since 
it demonstrates how sensitively the binding cooperativity depends 
on $\epsilon$.
In the regime of large $\epsilon$ the binding isotherm has a 
sharp (albeit continuous) slope at $\mu_{\rm s}^*$, as demonstrated 
by the dotted curve in fig.~\ref{results}a.
Hence, $\mu_{\rm s}=\mu_{\rm s}^*$ can be associated 
with $\phi_{\rm s}^*$,
the so-called {\em critical aggregation concentration (cac)} of 
surfactant, $\mu_{\rm s}^*=\ln\phi_{\rm s}^*$.

Our findings are summarized in the diagram drawn in fig.~\ref{results}b.
Depending on the value of $\epsilon$, two distinct binding regimes
are predicted. 
The non-cooperative binding regime of small $\epsilon$
($\alpha\gg\gamma$) corresponds to a situation where the surfactant 
almost does not affect the self-assembly of the polymeric side groups. 
The polymer is hydrophobic enough by itself to form microdomains
and the bound surfactants merely join the already existing 
micelles, swelling them a little.
On the other hand, the cooperative binding regime of large $\epsilon$
($\gamma\gg\alpha$) corresponds to the opposite situation 
where the surfactant triggers the self-assembly. 
Above the {\it cac}, the number of repeat units per hydrophobic
microdomain jumps from roughly $\alpha$ to about 
$\alpha+\gamma \gg \alpha$.

In recent experiments \cite{Zana}
both polymer hydrophobicity and surfactant-polymer affinity were
changed by controlling the polymer charge and salt concentration
(probably affecting both $\alpha$ and $\gamma$) and the length of the 
hydrophobic side chains (mainly affecting $\alpha$).
In those experiments the binding cooperativity could, indeed, 
be sensitively tuned by
changing either the polymer hydrophobicity or its affinity to the
surfactant, and the two distinct binding regimes described by our 
model were clearly observed.

Similar to regular surfactant micellisation, our self-assembly model 
yields a well-defined surfactant aggregation number
determined by the correlation length of the polymer chain 
(the number of repeat units per microdomain) and the average fraction 
of bound surfactant.
In our case of surfactant-polymer self-assembly, the
aggregate size is restricted by the entropy of the 
amphiphilic backbone, which imposes a finite correlation length
along the chain.
Moreover, like regular micellisation,
the self-assembly at the {\it cac}, though sharp, 
is not a first-order transition. 
In our model this is due to the one-dimensionality of 
the effective interactions between surfactants.

We have presented a model which 
accounts for the experimentally observed diverse behaviour of 
dilute solutions of amphiphilic side-chain polymers 
and surfactants.
Our results demonstrate the
delicate balance in these mixed systems.
By changing the parameters of the polymer and surfactant it is possible 
to cause aggregation and extensively modify the macroscopic properties
of the solution.
While capturing the thermodynamics of self-assembly
in the system and accounting for the polymer-induced interactions
between surfactants, our model still lacks the microscopic
structural details required for a more complete description of the
assembled microdomains themselves.
Such a refinement will be presented in a future paper \cite{future}.
Another important extension of the model is to consider
inter-chain association in more concentrated solutions and relate 
structural properties to rheological ones. 
\stars
We thank I. Borukhov, W. Gelbart, 
J.-F. Joanny, H. Orland, S. A. Safran, Z.-G. Wang and R. Zana for 
stimulating discussions.
DA would like to thank L. Leibler for introducing him to the subject of 
associating polymers and for many illuminating discussions.
Partial support from the Israel Science Foundation founded by
the Israel Academy of Sciences and Humanities -- Centers of 
Excellence Program -- and the U.S.-Israel Binational Foundation 
(B.S.F.) under grant no.~94-00291 is gratefully acknowledged.
%
%%%   references.
%

%
\end{document}